\documentclass[aip,cha,amsmath,amssymb,reprint]{revtex4-1}
\usepackage{graphicx}% Include figure files
\usepackage{dcolumn}% Align table columns on decimal point
\usepackage{bm}% bold math
\usepackage{hyperref}
\usepackage[rightcaption]{sidecap}
\usepackage{float}
\usepackage{amsmath}
\usepackage{courier}
\usepackage{braket}
\usepackage[usenames,dvipsnames]{xcolor}
\usepackage{upgreek}
\hypersetup{
  colorlinks   = true,	%Colours links instead of ugly boxes
  urlcolor     = blue,	%Colour for external hyperlinks
  linkcolor    = blue,	%Colour of internal links
  citecolor   = blue	%Colour of citations#NavyBlue
}
\newcommand{\ve}{\varepsilon}
\begin{document}
%\preprint{AIP/123-QED}

\title{Revival of oscillation and symmetry breaking in coupled quantum oscillators}

\author{Biswabibek Bandyopadhyay}
\affiliation{Chaos and Complex Systems Research Laboratory, Department of Physics, University of Burdwan, Burdwan 713
  104, West Bengal, India}
%\author{Taniya Khatun}
%\affiliation{Chaos and Complex Systems Research Laboratory, Department of Physics, University of Burdwan, Burdwan 713
%  104, West Bengal, India}
%\author{Debabrata Biswas}
%\affiliation{Department of Physics, Bankura University, Bankura 722 155, West Bengal, India}
\author{Tanmoy Banerjee}
\email[]{tbanerjee@phys.buruniv.ac.in}
%%%\homepage[]{Your web page}
%%%\thanks{}
%%%\altaffiliation{}
\affiliation{Chaos and Complex Systems Research Laboratory, Department of Physics, University of Burdwan, Burdwan 713
  104, West Bengal, India}
  
\received{}
\date{\today}

\begin{abstract}
Restoration of oscillation from an oscillation suppressed state in coupled oscillators is an important topic of research and has been studied widely in recent years. However, the same in the quantum regime has not been explored yet. Recent works established that under certain coupling conditions coupled quantum oscillators are susceptible to suppression of oscillations, such as amplitude death and oscillation death. In this paper, for the first time we demonstrate that quantum oscillation suppression states can be revoked and rhythmogenesis can be established in coupled quantum oscillators by controlling a feedback parameter in the coupling path. However, in sharp contrast to the classical system, we show that in the deep quantum regime the feedback parameter fails to revive oscillation, rather results in a transition from quantum amplitude death state to the recently discovered quantum oscillation death state. We use the formalism of open quantum system and phase space representation of quantum mechanics to establish our results. Therefore, our study establishes that revival scheme proposed for classical systems does not always result in restoration of oscillation in quantum systems but in the deep quantum regime it may give counterintuitive behaviors that are of pure quantum mechanical origin.            
\end{abstract}

\maketitle
\begin{quotation}
Understanding dissipative nonlinear systems in the quantum regime has recently been identified as an important topic of research. In the quantum domain the well known phenomena like synchronization and oscillation quenching show several significant results that have no counterpart in the classical domain. In this context, the restoration of oscillation from an oscillation suppressed state in coupled oscillators  has not been explored yet in the quantum regime. In this paper, for the first time we study the phenomenon of revoking quantum oscillation suppression states. We demonstrate that in the weak quantum regime the revival mechanism works properly. However, in sharp contrast to the classical system, we show that in the deep quantum regime, the process of rhythmogenesis fails, and results in a transition from quantum amplitude death state to quantum oscillation death state. We use the formalisms used in open quantum systems and noisy classical systems to establish our results.

\end{quotation}

\section{Introduction}
Exploring nonlinear dynamics in the open quantum systems has gained much attention in recent years \cite{lee_prl,brud_prl1,expt1,expt2}. The well known concepts of nonlinear dynamics such as oscillation of a single unit, and emergent behaviors of coupled oscillatory units, such as synchronization \cite{piko} have recently been explored in the quantum regime \cite{lee_prl,brud_prl1}. The extension of the techniques used in the so called classical nonlinear dynamics to the quantum regime is not always straightforward. Understanding of nonlinear behavior in the quantum domain is based on the formalism of open quantum system that requires the solution of quantum master equations \cite{carmichael}. Also, phase space representation of quantum system which involves quasi probability function (e.g. Wigner function \cite{wigner}) plays a crucial role in this endeavor.   

Nonlinear dynamics in the quantum regime is worth studying as the well known classical results deviate in the quantum regime and manifest several counterintuitive results that are not possible in the classical system. The notion of synchronization in the quantum regime is explored in detail both theoretically \cite{lee_prl,brud_prl1} and experimentally \cite{expt1,expt2}. 
Refs.~\onlinecite{lee_prl} and \onlinecite{brud_prl1} explored the role of the inherent quantum noise to defy synchronization in the quantum regime: Later on, several aspects of quantum synchronization \cite{lee_pre_1,brud-ann15,brudprl_17,morgan} and to improve the quality of synchronization \cite{squeezing,enhance-kwek} have been reported. The manifestation of partial synchronization or chimera state \cite{annabook,acamc,annapdeath} differs in the quantum regime\cite{schoell_qm}.  
In the context of control, the well known Pyragas control scheme \cite{py92} has been tested in the quantum regime \cite{schoellqm2}.
The phenomenon of coherence resonance has been found to deviate in the quantum regime and shows improvement of regular behavior in the presence of inherent quantum noise \cite{qmcores}
Subsequently, quantum manifestation of another widely studied emergent dynamics, namely oscillation quenching has also been studied recently. Ishibashi \textit{et. al.} \cite{qad1} demonstrated that, similar to classical system, in quantum oscillators also parameter mismatch in diffusive coupling leads to amplitude death (AD) where all the oscillators arrive at the common steady state. However, unlike classical system, complete suppression of oscillation is restricted by the inherent quantum noise present in the quantum oscillators. Nevertheless, a pronounced decrease in the mean boson number is found to be indicative of quantum AD. Later, Amitai \textit{et. al.} \cite{qad2} showed that the Kerr-type anharmonicity is conducive to quantum AD. Apart from AD there exists a symmetry-breaking version of oscillation suppression called oscillation death (OD) in which coupling dependent nontrivial steady states are created \cite{kosprep,kosprl}; recently \citet{qmod} discovered the quantum analogue of the OD state.

In this context an important emergent dynamics, namely the revival of oscillation from the oscillation suppressed state has yet not been explored in the quantum regime. Revival of oscillation is important in several physical and biological systems as maintaining rhythmicity is often desirable in those systems. Examples include power grid, El nino, sinuartial rhythm where cessation of oscillations may cause serious consequences. Several techniques of revival of oscillations have been proposed in the literature all of which are based on either controlling the delay or the dissipation rate of the coupling function. The most general technique of revival of oscillation has been reported by Zou \textit{et. al.} \cite{natcom} that is based on the introduction of a feedback parameter in the coupling path. This technique modifies the dissipation rate in the coupling path and works successfully over a broad range of coupling functions and oscillators \cite{tanryth}. However, all these revival schemes are studied in the classical regime.

In this paper, for the first time, we study the revival of oscillations in the hitherto unexplored quantum regime. Our main aim in this paper is to test the applicability and efficacy of the ``dissipation-modified"  revival scheme of Ref.~\onlinecite {natcom,tanryth}. For our study we consider the paradigmatic  quantum van der Pol oscillators with a coupling scheme, which is known to induce ``death", namely the mean-field diffusive coupling \cite{tanCD,bandutta}. We show that revival of oscillation from quantum amplitude death state is indeed possible in the weak quantum regime. The revival of oscillation is manifested by the pronounced increase in the mean phonon number. However, in sharp contrast to the classical and weak quantum case, in the deep quantum regime the technique does not result in revival of oscillation, rather the system is trapped in either homogeneious or inhomogeneous quantum steady states; the variation of the feedback parameter results in a transition from quantum AD state to the quantum OD state that has no counterpart in the semiclassical regime. The present study asserts that the revival of oscillations manifests results that are exclusive to quantum domain only.
%facilitate a quantum Turing like transition from QAD to QOD.    

The rest of the paper is organized as follows: the next section provides the mathematical model and manifestation of oscillation of a single quantum van der Pol oscillator. Section~\ref{sec:mf} presents the classical revival scheme in coupled oscillators under mean-field diffusive coupling. Section~\ref{sec:qm} describes the revival and symmetry breaking phenomena in coupled quantum oscillators; it formulates the quantum master equation and the results of weak and deep quantum quantum regime. In the weak quantum regime a semiclassical treatment with noisy classical model is also presented in this section. Finally, we summarize the results in Sec.~\ref{sec:con}.      
\begin{figure}
%%\vspace{0.4cm}
\includegraphics[width=.48\textwidth]{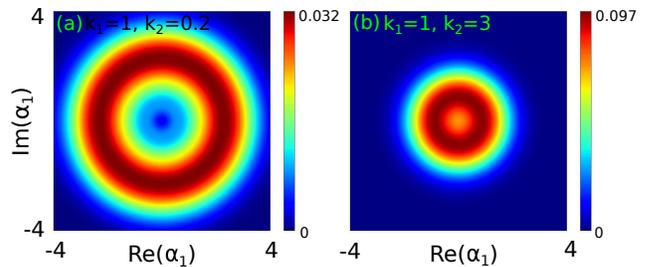}
\caption{Limit cycle from quantum van der Pol oscillator (Eq.~\ref{single_master}). (a) Weak quantum regime: $k_1=1$ and $k_2=0.2$. (b) Deep quantum regime: $k_1=1$, and $k_2=3$. ($\omega=2$).}
\label{phase_space}
\end{figure}

%##############################################################
%#####################Revival Scheme###########################
%##############################################################

\section{Quantum van der Pol oscillator}
\label{sec:vdp}
%==========================(xy equation)==============================
Let us at first describe a single van der Pol oscillator in the classical and quantum domain. A {\it classical} van der Pol oscillator has the following mathematical form \cite{vdp,brud_prl1}:
\begin{subequations}
\label{vdp}
\begin{align}
\dot{x}&=\omega y, \\
%\end{align}
%\begin{align}
\dot{y}&=-\omega x+(k_1-8k_2 {x}^2)y,
\end{align}
\end{subequations}
%\begin{equation}
%\label{vdp}
%\ddot{x}=-\omega^2x+k_1\dot{x}-8k_2x^2\dot{x},
%\end{equation}
where $\omega$ is the eigenfrequency of the oscillator. $k_1$ and $k_2$ control the linear gain and the nonlinear damping, respectively ($k_1,k_2>0$).
%==========================(Amplitude equation)==============================
By using harmonic approximations\cite{qmod} we derive the following amplitude equation of Eq.~\eqref{vdp} in terms of a complex amplitude, $\alpha=x+iy$: 
\begin{equation}
\label{single_amp}
\dot{\alpha}=-i\omega\alpha+(\frac{k_1}{2}-k_2|\alpha|^2)\alpha.
\end{equation}
The oscillator shows a limit cycle oscillation with an amplitude $\sqrt{\frac{k_1}{2k_2}}$.

%==========================(Master equation)==============================
Let us write the quantum master equation corresponding to the {\it quantum} van der Pol oscillator \cite{lee_prl,brud_prl1}:
\begin{equation}
\label{single_master}
\dot{\rho}=-i[\omega a^\dag a,\rho]+k_1\mathcal{D}[a^\dag](\rho)+k_2\mathcal{D}[a^2](\rho).
\end{equation}
This equation actually represents a harmonic oscillator interacting with the environment (bath) through the Lindblad dissipator $\mathcal{D}[\hat{O}](\rho)$: where $\mathcal{D}[\hat{O}](\rho)=\hat{O}\rho \hat{O}^\dag-\frac{1}{2}\{\hat{O}^\dag \hat{O},\rho \}$ (we consider $\hbar=1$). $a$ and $a^\dag$ are the bosonic annihilation and creation operators, respectively. $k_1$ is the linear  pumping rate that controls the creation of a single boson and $k_2$ represents the nonlinear loss rate governing the annihilation of two bosons. Eq.~\ref{single_master} has been widely studied in the context of synchronization and oscillation suppression. In the classical limit (or weak quantum limit), linear pumping ($k_1$) dominates over the nonlinear loss (i.e., $k_1>k_2$) and in this higher excitation region one can consider $\langle a\rangle \equiv \alpha$, and the master equation \eqref{single_master} and the classical amplitude equation \eqref{single_amp} are equivalent through the following relation: $\dot{\braket{a}}=\mbox{Tr}(\dot \rho a)$. Figure~\ref{phase_space}(a) shows the steady state Wigner function based phase space representation of the quantum limit cycle (using \texttt{QuTiP} \cite{qutip}) in the weak quantum limit ($k_1=1$ and $k_2=0.2$). In the limit where $k_2>k_1$ the system resides in the deep quantum regime. In this limit only a few Fock levels near the quantum mechanical ground state are populated: Fig.~\ref{phase_space}(b) shows the corresponding limit cycle for $k_1=1$ and $k_2=3$. In the limiting case of $k_2\rightarrow \infty$ the steady state density matrix is \cite{lee_prl} $\rho_{ss}=\frac{2}{3}|0\rangle\langle 0|+\frac{1}{3}|1\rangle\langle 1|$. i.e., the oscillation still persists unlike classical vdP oscillator.

%####################### (Section) ############################
%##############################################################

\section{Classical Revival of oscillation}\label{sec:mf} 

%===============================================================
%======================= (Sub section) =========================
%===============================================================

Before proceed to the quantum case, let us demonstrate the revival scheme of two identical classical van der Pol oscillators that are coupled via weighted mean-field diffusive coupling. A mathematical model is given by\cite{qmod}
%~~~~~~~~~~~~~~~~~~~~~~~~~~~~~~~~~~~~~~~~~~~~b_eqn
\begin{subequations}
\label{scalar_classical}
\begin{align}
\dot{x}_j&=\omega y_j+\varepsilon\left(\frac{q}{2}\sum_{m=1}^2x_m-\upgamma x_j\right), \\
%\end{align}
%\begin{align}
\dot{y}_j&=-\omega x_j+(k_1-8k_2 {x_j}^2)y_j,
\end{align}
\end{subequations}
%~~~~~~~~~~~~~~~~~~~~~~~~~~~~~~~~~~~~~~~~~~~~e_eqn
$j\in\{1, 2\}$. $\varepsilon$ is the coupling parameter. Both the oscillators have the common eigenfrequency $\omega$. The control parameter $q$ determines the density of the weighted mean-field, which is relevant in several processes including quantum physics \cite{qmod} and biology \cite{qstr2}. Here the parameter $\upgamma$ controls the dissipation rate in the coupling part: it actually controls the asymmetry between the incoming and outgoing flow of the diffusion process\cite{natcom}. $\upgamma=1$ represents the normal mean-field diffusive coupling. By decreasing $\upgamma$ below unity ($0<\upgamma<1$) one reduces the dissipation rate in the coupling channel which is conducive for rhythmic behavior \cite{natcom,tanryth}. The introduction of $\upgamma$ is especially relevant in the {\it open quantum systems} as here the dynamics is essentially dissipative.

The system Eq.~\ref{scalar_classical} has two types of fixed points, namely trivial fixed point $\mathcal{F}_{HSS} \equiv (0, 0, 0, 0)$ and nontrivial fixed point $\mathcal{F}_{IHSS}=(x^*,y^*,-x^*,-y^*)$, where $x^*=\frac{\omega y^*}{\varepsilon \upgamma}$ and $y^*=\pm\sqrt{\frac{1}{8k_2}\left(\frac{k_1\varepsilon^2\upgamma^2}{\omega^2}-\varepsilon\upgamma\right)}$. The system shows a transition from oscillatory state to amplitude death state through an inverse Hopf bifurcation at $\varepsilon_{\mbox{HB}}=\frac{k_1}{(\upgamma-q)}$ \cite{tanpre1,tanpre2}. The classical equation Eq.~\eqref{scalar_classical} shows a transition from AD to OD state through a  pitchfork bifurcation at $\epsilon_{\mbox{PB}}=\frac{\omega^2}{k_1 \upgamma}$ \cite{tanpre1,tanpre2}. The Hopf and pitchfork bifurcation curves in the $\ve/k_1-\upgamma$ space are shown in Fig.~\ref{s_twopar} ($q=0.2$). In the classical model above the Hopf bifurcation curve (HB) amplitude death occurs. However, if one decreases the control parameter $\upgamma$, below the HB curve the death state is revoked and oscillatory behavior is restored. Therefore, the HB curve is the oscillation revival curve. In the next section we discuss how this revival scenario holds in quantum systems.     
%===============================================================
%======================= (Sub section) =========================
%===============================================================

\section{Pure Quantum Model: Revival and symmetry breaking}\label{sec:qm}

\subsection{quantum master equation}
The quantum master equation of two scalar mean-field diffusively coupled identical quantum van der Pol oscillators under revival scheme is given by,
%~~~~~~~~~~~~~~~~~~~~~~~~~~~~~~~~~~~~~~~~~~~~b_eqn
\begin{equation}
\label{master_scalar}
\begin{split}
\dot{\rho}&=-i\bigg[\omega (a_1^\dag a_1+a_2^\dag a_2)+\frac{i\epsilon}{4}\big(q(a_1^\dag a_2^\dag-a_{1}a_{2})\\
&+(\frac{q}{2}-\upgamma)({a_1^\dag}^2+{a_2^\dag}^2-{a_1}^2-{a_2}^2)\big),\rho\bigg]\\
&+k_1\sum_{j=1}^2\mathcal{D}[a_j^\dag](\rho)+k_2\sum_{j=1}^2\mathcal{D}[{a_j}^2](\rho)\\
&+\frac{q\epsilon}{2}\mathcal{D}[(a_1+a_2)^\dag](\rho)+\epsilon\upgamma\sum_{j=1}^2\mathcal{D}[a_j](\rho).
\end{split}
\end{equation}
%~~~~~~~~~~~~~~~~~~~~~~~~~~~~~~~~~~~~~~~~~~~~e_eqn
where $a_j$ ($a_j^\dag$) is the annihilation (creation) operator corresponding to the $j$-th oscillator.  
%%%%%%Correspondence between Master equation and Amplitude equation}
In the classical limit ($k_1>k_2$), using the relation $\dot{\braket{a}}=\mbox{Tr}(\dot \rho a)$, the master equation \eqref{master_scalar} is equivalent to the classical amplitude equation 
of Eq.~\eqref{scalar_classical}, which reads
%~~~~~~~~~~~~~~~~~~~~~~~~~~~~~~~~~~~~~~~~~~~~b_eqn
\begin{equation}
\label{amp_eqn_scalar}
\begin{split}
\dot{\alpha_j}&=-i\omega\alpha_j+(\frac{k_1}{2}-k_2|\alpha_j|^2)\alpha_j+\frac{\varepsilon}{2}\left(\frac{q}{2}\sum_{m=1}^{2}\alpha_m-\upgamma\alpha_j\right)\\
&+\frac{\varepsilon}{2}\left(\frac{q}{2}\sum_{m=1}^{2}\alpha^*_m-\upgamma\alpha^*_j\right).
\end{split}
\end{equation}

%=====================bf
\begin{figure}
%\vspace{0.4cm}
\includegraphics[width=.48\textwidth]{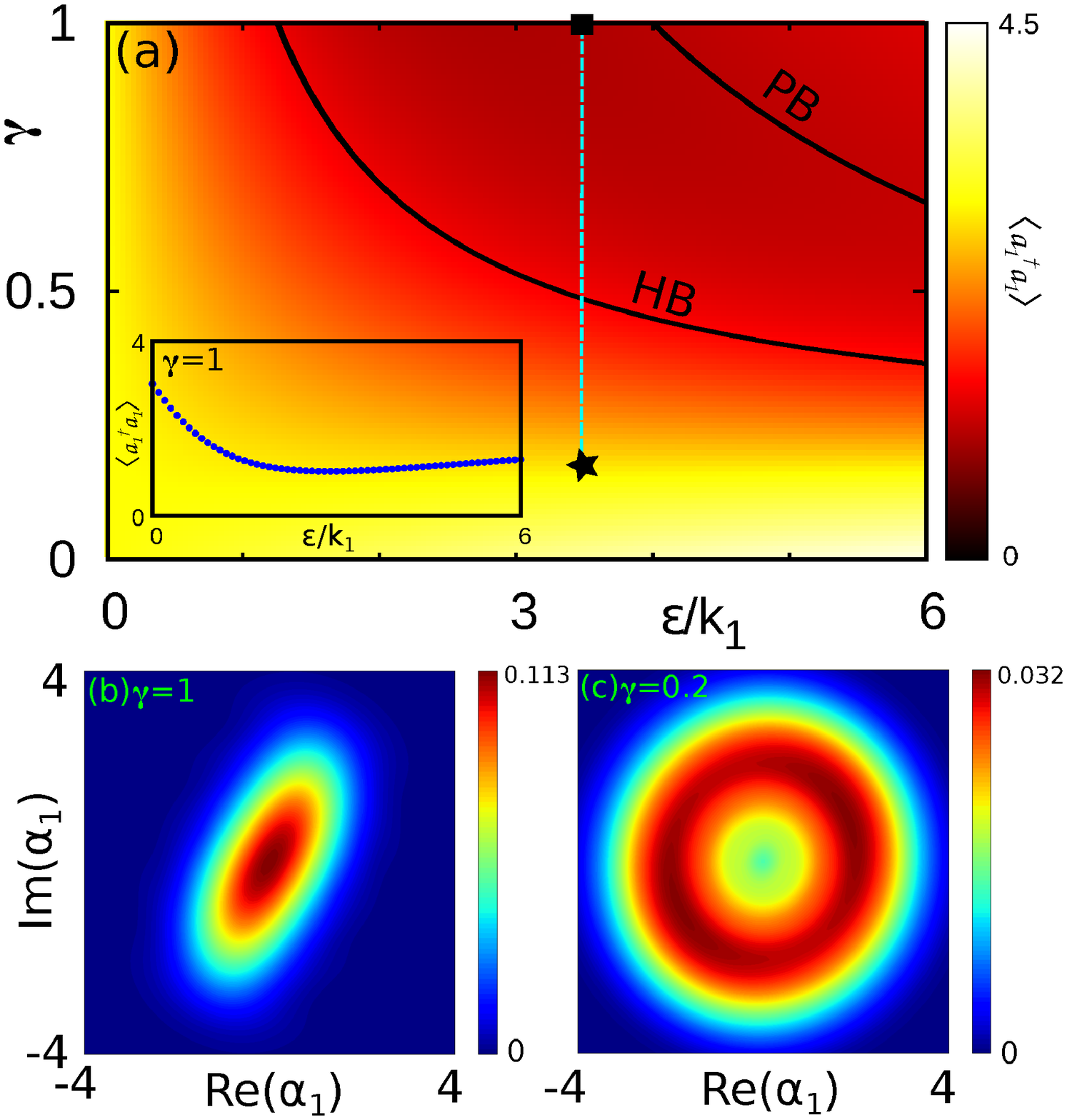}
\caption{{\bf Weak quantum region:} (a) Colormap of mean-phonon number of the first oscillator ($\braket{a_{1}^\dag a_{1}}$) in the $\ve/k_1-\upgamma$ space at $q=0.2$. The solid lines represent Hopf bifurcation curve (HB) and pitchfork bifurcation curve (PB) drawn using $\varepsilon_{\mbox{HB}}$ and $\varepsilon_{\mbox{PB}}$ (Sec.~\ref{sec:mf}). Two points at the same $\ve/k_1$ ($\ve/k_1=3.5$) and different $\upgamma$ (at solid square $\upgamma=1$ and at star $\upgamma=0.2$) are marked on either side of the classical Hopf bifurcation curve. The inset shows $\braket{a_{1}^\dag a_{1}}$) vs. $\ve/k_1$ curve for $\upgamma=1$. The Wigner function distributions in phase space corresponding to these two points are shown in (b) $\upgamma=1$: quantum AD and (c) $\upgamma=0.2$: quantum limit cycle. Other parameters are ($k_1, k_2) = (1, 0.2$), and $\omega=2$.}
\label{s_twopar}
\end{figure}

\subsection{Weak quantum regime: revival of oscillations}

\subsubsection{Quantum results}
At first, we consider the weak quantum regime where excitations are moderate or large. We ensure this region by taking the pumping rate ($k_1$) higher than the damping rate ($k_2$). We solve the master equation \eqref{master_scalar} numerically using \texttt{QuTiP} \cite{qutip} for the following parameters: $k_1=1$ and $k_2=0.2$. 
Fig.~\ref{s_twopar}~(a) shows the color map of the mean phonon number of the first oscillator ($\braket{a_{1}^\dag a_{1}}$) in the $\ve/k_1-\upgamma$ space at $q=0.2$. It can be seen from the figure that for $\upgamma=1$ quantum amplitude death (QAD) occurs for $\ve/k_1>\ve_{\mbox{HB}}/k_1$. This is manifested by the pronounced decrease of mean phonon number ($\braket{a_{1}^\dag a_{1}}$) with increasing $\ve/k_1$ (see the inset of Fig.~\ref{s_twopar}~(a)). The corresponding Wigner function is shown in Fig.~\ref{s_twopar}~(b) that depicts a {\it squeezed} quantum AD state for $\upgamma=1$ and $\ve/k_1=3.5$. Compare it with the limit cycle of Fig.~1(a) (for $\upgamma=1$ and $\ve/k_1=0$) where the Wigner function shows maximum distribution at values far from the origin, however, in the quantum AD state the Wigner function shows maximum probability around the origin indicating sharp decrease in amplitude. However, the inherent quantum noise resists the complete cessation of amplitude. 

This quantum AD state can be revoked by decreasing the dissipation parameter $\upgamma$. We find that by decreasing $\upgamma$ value from a point above the HB curve (i.e., quantum AD state) we can enter into the quantum oscillatory region (below the HB curve) at a particular value of $\ve/k_1$. From Fig.~\ref{s_twopar}~(a) it can be seen that decrease in $\upgamma$ below the HB curve results in the pronounced increase in the mean phonon number. This is demonstrated for an exemplary value of $\upgamma=0.2$ and $\ve/k_1=3.5$ in Fig.~\ref{s_twopar}~(c). In contrast to Fig.~\ref{s_twopar}~(b) ($\upgamma=1$), Fig.~\ref{s_twopar}~(c) shows a Wigner distribution that resides around the non-zero values indicating the occurrence of quantum limit cycle. Therefore, $\upgamma$ acts as a control parameter and responsible for the occurrence of revival of quantum oscillation. However, in agreement with Ref.~\onlinecite{qmod} we do not get any quantum OD state in the weak quantum regime even in the presence of the dissipation parameter $\upgamma$. Therefore, the classical PB curve does not play any role in the weak quantum region.

%===============================================================
%======================= (Sub section) =========================
%===============================================================

\subsubsection{Noisy classical results}

For better understanding of the revival phenomenon we study the noisy classical model and compare the results with that of quantum systems. In both the cases equal noise intensity has to be considered. For this, the quantum master equation \eqref{master_scalar} is represented in phase space using partial differential equation of Wigner distribution function \cite{carmichael}.
%~~~~~~~~~~~~~~~~~~~~~~~~~~~~~~~~~~~~~~~~~~~~b_eqn
\begin{equation}
\label{diff_eqn_w_s}
\begin{split}
\dot{W}&=\sum_{j=1}^2\left[-\left(\frac{\partial}{\partial \alpha _j}\mu _{\alpha _j}+c.c.\right) \right. \\
&+ \left. \frac{1}{2}\left(\frac{\partial ^2}{\partial \alpha _j \partial {\alpha_j}^*}D_{\alpha _j {\alpha_j}^*}+\frac{\partial ^2}{\partial \alpha _j \partial {\alpha_{j'}}^*}D_{\alpha _j {\alpha_{j'}}^*} \right) \right. \\
&+ \left. \frac{k_2}{4}\left(\frac{\partial ^3}{\partial {\alpha_j}^* \partial {\alpha_j}^2}\alpha _j+c.c\right) \right]W,
\end{split}
\end{equation}
%~~~~~~~~~~~~~~~~~~~~~~~~~~~~~~~~~~~~~~~~~~~~e_eqn
where $\mu$ and $D$ are respectively the elements of drift vector and diffusion matrix. Here, $\mu _{\alpha _j}=\left[-i\omega+\frac{k_1}{2}-k_2(|\alpha _j|^2-1)\right]\alpha _j-\left(\frac{\varepsilon\upgamma}{2}-\frac{\varepsilon q}{4}\right)(\alpha _j + {\alpha _j}^*)+\frac{\varepsilon q}{4}(\alpha _{j'}+{\alpha _{j'}}^*)$, $D_{\alpha _j {\alpha_j}^*}=k_1+2k_2(2|\alpha _j|^2-1)+\frac{\varepsilon q}{2}+\varepsilon\upgamma$ and $D_{\alpha _j {\alpha_{j'}}^*}=\frac{\varepsilon q}{2}$ with $j=1,2$, $j'=1,2$ and $j\neq j'$. In weak nonlinear regime ($k_2<k_1$), the third-order terms can be ignored and this differential equation is reduced to the Fokker-Planck equation, which is given below for our system:
%~~~~~~~~~~~~~~~~~~~~~~~~~~~~~~~~~~~~~~~~~~~~b_eqn
\begin{equation}
\label{fp_s}
\begin{split}
\dot{W}(\textbf{X})&=\sum_{j=1}^2\left[-\left(\frac{\partial}{\partial x_j}\mu _{x_j}+\frac{\partial}{\partial y_j}\mu _{y_j}\right) \right. \\
&+ \left. \frac{1}{2}\left(\frac{\partial ^2}{\partial x_j \partial x_j}D_{x_j x_j}+\frac{\partial ^2}{\partial y_j \partial y_j}D_{y_j y_j} \right. \right. \\
&+ \left. \left. \frac{\partial ^2}{\partial x_j \partial x_{j'}}D_{x_j x_{j'}}+\frac{\partial ^2}{\partial y_j \partial y_{j'}}D_{y_j y_{j'}}  \right) \right]W(\textbf{X}),
\end{split}
\end{equation}
%~~~~~~~~~~~~~~~~~~~~~~~~~~~~~~~~~~~~~~~~~~~~e_eqn
where $\textbf{X}=(x_1, y_1, x_2, y_2)$. The elements of drift vector are,
%~~~~~~~~~~~~~~~~~~~~~~~~~~~~~~~~~~~~~~~~~~~~b_eqn
\begin{subequations}
\label{drift_s}
\begin{align}
\begin{split}
\mu _{x_j}&=\omega y_j + \left[\frac{k_1}{2}-k_2({x_j}^2+{y_j}^2-1) \right. \\
&- \left. \left(\varepsilon\upgamma-\frac{\varepsilon q}{2}\right)\right]x_j+\frac{\varepsilon q}{2}x_{j'},
\end{split} \\
\begin{split}
\mu _{y_j}&=-\omega x_j + \left[\frac{k_1}{2}-k_2({x_j}^2+{y_j}^2-1)\right]y_j. 
\end{split}
\end{align}
\end{subequations}
%~~~~~~~~~~~~~~~~~~~~~~~~~~~~~~~~~~~~~~~~~~~~e_eqn

The diffusion matrix has the following form,
%~~~~~~~~~~~~~~~~~~~~~~~~~~~~~~~~~~~~~~~~~~~~b_eqn
\begin{align}\label{diffusion_mat_s} \nonumber
{\mbox {\bf D}}&=\left(\begin{array}{cccc} D_{x_1 x_1} & 0 & D_{x_1 x_2} & 0 \\
0 & D_{y_1 y_1} & 0 & D_{y_1 y_2}\\
D_{x_2 x_1} & 0 & D_{x_2 x_2} & 0 \\
 0 & D_{y_2 y_1} & 0 & D_{y_2 y_2}  \end{array}\right) \\ 
&=\frac{1}{2}\left(\begin{array}{cccc} \nu _1 & 0 & \frac{\varepsilon q}{4} & 0 \\
0 & \nu _1 & 0 & \frac{\varepsilon q}{4}\\
\frac{\varepsilon q}{4} & 0 & \nu _2 & 0 \\
 0 & \frac{\varepsilon q}{4} & 0 & \nu _2  \end{array}\right),
\end{align}
%~~~~~~~~~~~~~~~~~~~~~~~~~~~~~~~~~~~~~~~~~~~~e_eqn
where $\nu _j=\frac{k_1}{2}+k_2[2({x_j}^2+{y_j}^2)-1]+\frac{\varepsilon q}{4}+\frac{\varepsilon\upgamma}{2}$.

From Eq.\eqref{fp_s} the following stochastic differential equation can be derived:
%~~~~~~~~~~~~~~~~~~~~~~~~~~~~~~~~~~~~~~~~~~~~b_eqn
%\begin{subequations}
\begin{equation}\label{sde_s}
d\textbf{X}=\bm{\mu}dt+\bm{\sigma} d\textbf{W}_t,
\end{equation}
%\end{subequations}
%~~~~~~~~~~~~~~~~~~~~~~~~~~~~~~~~~~~~~~~~~~~~e_eqn
where $\bm{\sigma}$ is the noise strength and $d\textbf{W}_t$ is the Wiener increment. As the diffusion matrix $\textbf{D}$ (given in Eq.\eqref{diffusion_mat_s}) is symmetric, we can analytically derive $\bm{\sigma}$ from it. First, $\textbf{D}$ has to be diagonalized. The diagonal form of $\textbf{D}$ can be written as $\textbf{D}_{diag}=\textbf{U}^{-1}\textbf{D}\textbf{U}=diag(\lambda_- \lambda_- \lambda_+ \lambda_+)$. Where $\lambda_{\pm}=\frac{1}{4}\left[\nu _1+\nu _2\pm\sqrt{(\nu _1-\nu _2)^2+(\frac{\varepsilon q}{2})^2}\right]$ and $\textbf{U}$ has the following form:
%~~~~~~~~~~~~~~~~~~~~~~~~~~~~~~~~~~~~~~~~~~~~b_eqn
\begin{align}\label{u_mat_s}
{\mbox {\bf U}}&=\left(\begin{array}{cccc} 0 & u_- & 0 & u_+ \\
u_- & 0 & u_+ & 0\\
0 & 1 & 0 & 1 \\
1 & 0 & 1 & 0  \end{array}\right),
\end{align}
%~~~~~~~~~~~~~~~~~~~~~~~~~~~~~~~~~~~~~~~~~~~~e_eqn
where $u_{\pm}=\frac{2}{\varepsilon q}\left[\nu _1-\nu _2\pm\sqrt{(\nu _1-\nu _2)^2+(\frac{\varepsilon q}{2})^2}\right]$. Now, $\bm{\sigma}$ matrix can be evaluated from the equation $\bm{\sigma}=\textbf{U}\sqrt{\textbf{D}_{diag}}\textbf{U}^{-1}$ and it has the following form:
%~~~~~~~~~~~~~~~~~~~~~~~~~~~~~~~~~~~~~~~~~~~~b_eqn
\begin{align}\label{sigma_mat_s}
\bm{\sigma}&=\left(\begin{array}{cccc} \sigma_1 & 0 & \sigma_3 & 0 \\
0 & \sigma_1 & 0 & \sigma_3\\
\sigma_3 & 0 & \sigma_2 & 0 \\
0 & \sigma_3 & 0 & \sigma_2  \end{array}\right),
\end{align}
%~~~~~~~~~~~~~~~~~~~~~~~~~~~~~~~~~~~~~~~~~~~~e_eqn
where $\sigma_1=\frac{u_+\sqrt{\lambda_+}-u_-\sqrt{\lambda_-}}{u_+-u_-}$, $\sigma_2=\frac{u_+\sqrt{\lambda_-}-u_-\sqrt{\lambda_+}}{u_+-u_-}$ and $\sigma_3=\frac{\sqrt{\lambda_+}-\sqrt{\lambda_-}}{u_+-u_-}$.

By solving the stochastic differential equation (Eq.~\eqref{sde_s}) (using JiTCSDE module in Python \cite{jitcode}) we compute the ensemble average of the squared steady-state amplitude of the first oscillator ($\overline{{|\alpha_1|_{nc}}^2}$), averaged over 1000 realizations, starting from random initial conditions.

%=====================bf
\begin{figure}
\includegraphics[width=.45\textwidth]{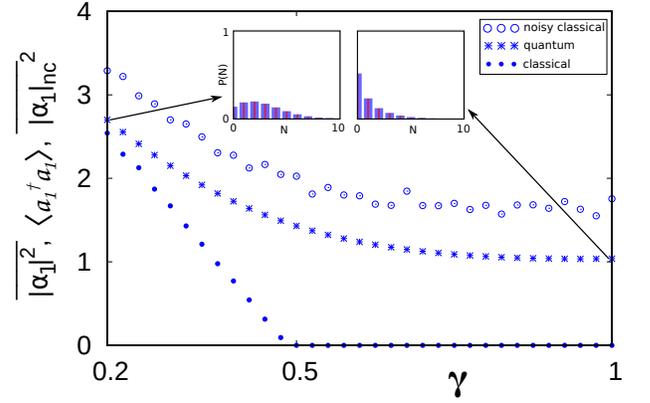}
\caption{Comparison of the classical, quantum, and noisy classical results with varying $\upgamma$ at $\ve/k_1=3.5$ and $q=0.2$ (i.e., along the vertical line joining two representative points in Fig.~\ref{s_twopar}(a)). the average amplitude from the classical model ($\overline{{|\alpha _1|}^2}$), mean phonon number from the quantum model ($\braket{a_1^\dag a_1}$) and the averaged amplitude from the noisy classical model ($\overline{{|\alpha _1|_{nc}}^2}$) of the first oscillator plotted together with $\upgamma$. The insets show occupation of Fock levels for the quantum amplitude death state $\upgamma=1$ (right panel) and the quantum oscillatory state $\upgamma=0.2$ (left panel) and . Other parameters are: $k_1=1$, $k_2=0.2$, and $\omega=2$.}
\label{s_1p_allmodels}
\end{figure} 
%=====================ef

Fig.~\ref{s_1p_allmodels} shows the plots of average amplitude and mean phonon number in classical, noisy classical and quantum oscillators with the variation of $\upgamma$ at $\ve/k_1=3.5$ and $q = 0.2$ (along the vertical line of Fig.~2(a)). At this value, $\upgamma=1$ gives amplitude death in all the three cases.  The classical AD state shows zero amplitude, i.e, $\overline{{|\alpha _1|}^2}=0$. However, for quantum and noisy classical cases the inherent noise resists the complete cessation of oscillation. The occupation of Fock levels at the quantum AD state is shown in the inset (right) for $\upgamma=1$ indicating the fact that the ground state is the most populated state. However, with decreasing $\upgamma$, average amplitude and mean phonon number increase indicating revival of oscillation in classical, noisy classical and quantum systems. However, unlike the classical case no sharp transition from amplitude death state to oscillatory state is possible in quantum and noisy classical cases due to the presence of the inherent noise. The occupation of Fock levels at the revived quantum oscillatory state is shown in the inset (left) for $\upgamma=0.2$ showing that the ground state is now depleted of phonon.  

%=====================bf
\begin{figure}
\includegraphics[width=.46\textwidth]{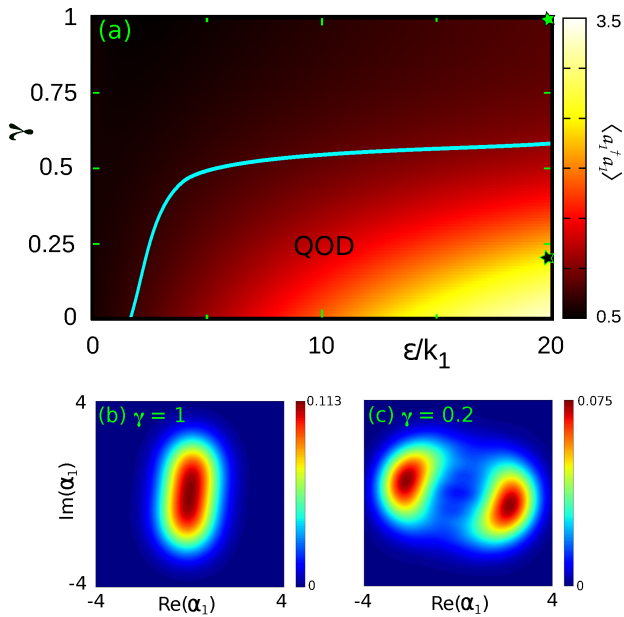}
\caption{{\bf Deep quantum regime:} In the deep quantum region ($k_1=1$ and $k_2=3$) two-parameter plot of mean phonon number ($\braket{a_1^\dag a_1}$) in the $\upgamma-\ve/k_1$ space; The quantum OD state appears below the cyan curve. The Wigner function distribution in phase space at (b) $\upgamma=1$ showing quantum AD and (c) $\upgamma=0.2$ showing quantum OD. Other parameters are $q=0.6$, $\ve/k_1=20$ and $\omega=2$.}
\label{s_k2_3}
\end{figure} 
%=====================ef
%===============================================================
%======================= (Sub section) =========================
%===============================================================

\subsection{Deep quantum regime: symmetry-breaking death state}
\begin{figure}
\centering
\includegraphics[width=0.42\textwidth]{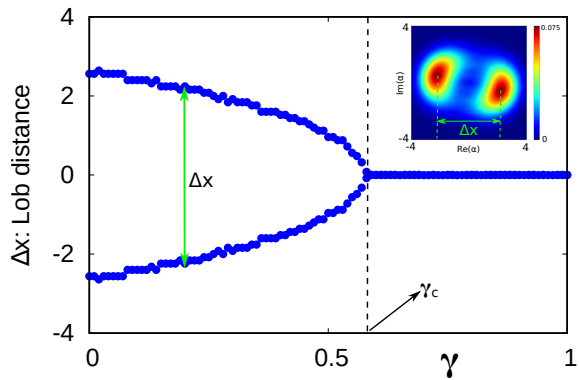}
\caption{{\bf Deep quantum regime:} Bifurcation diagram showing the displacement of the maximum values of Wigner function's lob(s) projected on the $x$-axis [i.e., Re($\alpha_1$)] denoted as $\Delta X$ with the variation of revival parameter $\upgamma$. Other parameters are $q=0.6$, $\varepsilon/k_1=20$, $k_2/k_1=3$, and $\omega=2$.}
\label{xdis}
\end{figure}

Finally, we study the effect of revival scheme in the deep quantum regime ($k_2>k_1$). Surprisingly, in stark contrast to the classical case, in the deep quantum regime a decrease in the dissipation parameter $\upgamma$ does not result in the revival of oscillation, rather it leads to an interesting symmetry-breaking steady state, namely quantum oscillation death (OD) state that is manifested in the phase space through a bimodal Wigner function. This state has recently been discovered in Ref.~\onlinecite{qmod} in the deep quantum regime under mean-field diffusive coupling. In the bimodal Wigner function the two lobes are equivalent to the two branches of the pitchfork bifurcation of classical case representing inhomogeneous steady states. Figure~\ref{s_k2_3}(a) shows the mean phonon number ($\braket{a_1^\dag a_1}$) in the $\upgamma-\ve/k_1$ space. Above the cyan curve unimodal Wigner function appears representing either quantum limit cycle or quantum AD. The quantum OD state appears below the cyan curve where the Wigner function becomes bimodal. Figure~\ref{s_k2_3}(b) shows the Wigner function in phase space at $\upgamma=1$ which is unimodal in nature representing quantum AD and Fig.~\ref{s_k2_3}(c) shows the same but for $\upgamma=0.2$ showing bimodal Wigner function representing the occurrence of quantum OD. We propose a numerical measure to mark the occurrence of quantum OD, which is based on the distance between the maximum values of the lobes of the Wigner function projected on the horizontal axis ($\Delta X$) [see the inset of Fig.~\ref{xdis}]. In the case of unimodal Wigner function  $\Delta X=0$. However, for the quantum OD, the Wigner function becomes bimodal, thus giving rise to two lobes separated by $\Delta X\ne 0$. In Fig.~\ref{xdis} we plot  $\Delta X$ with $\upgamma$. It shows that in the range $\upgamma_c \le \upgamma \le 1$ the Wigner function is unimodal. However, for $\upgamma<\upgamma_c$, with decreasing $\upgamma$  bimodal Wigner function appears having $\Delta X\ne 0$ mimicking symmetry-breaking classical pitchfork bifurcation and the appearance of quantum OD. However, it should be noted that in the deep quantum regime there is no one to one correspondence between the quantum and classical results.

%##############################################################
%####################### (Section) ############################
%##############################################################

\section{Conclusions}
\label{sec:con}
In this paper we have studied the revival of oscillation in coupled quantum oscillators by tuning a feedback parameter that controls the rate of dissipation in the coupling path. In coupled classical oscillators this was shown to be a general technique to revive oscillation from a oscillation suppressed state \cite{natcom,tanryth}. However, we have shown that although this technique works in revival of oscillation in the weak quantum regime, in stark contrast to the classical and semiclassical results, in the deep quantum regime it fails to do so; rather in this regime it leads to a symmetry-breaking transition from homogeneous oscillation quenching state (i.e., quantum amplitude death) to inhomogeneous oscillation quenching state (i.e., quantum oscillation death state). Our results established that, unlike classical systems, controlling the dissipation rate of the coupling path is not sufficient for the revival of oscillation in the deep quantum regime.  

In this paper we studied the ``dissipation-modified'' revival of oscillations in the quantum regime. However, we believe to get similar results in the case of other revival  schemes, such as  introduction of local low-pass filter \cite{lpf,banihlc} or time delay in the coupling path \cite{dvprl,tbook}, as in those cases also dissipation rate is being modified.

Apart from the academic interest, the revoking of a death state may play a crucial role in the quantum nanomechanical \cite{21_bemani,22_shim} and optomechanical \cite{expt-mem} oscillators where maintaining rhythmicity is crucial for proper functioning of the systems. Also, our study will be helpful in creating the squeezed and symmetry-breaking squeezed states that are found to be useful in several physical applications and precision measurements \cite{sq-app2,sq-app3}.

\begin{acknowledgments}
B.B. acknowledges the financial assistance from the University Grants Commission (UGC), India. T. B. acknowledges the financial support from the Science and Engineering Research Board (SERB), Government of India, in the form of a Core Research Grant [CRG/2019/002632].
\end{acknowledgments}

\section*{Data Availability Statement}
The data that support the findings of this study are available from the corresponding author upon reasonable request.

%\bibliography{qrev}

\begin{thebibliography}{44}%
\makeatletter
\providecommand \@ifxundefined [1]{%
 \@ifx{#1\undefined}
}%
\providecommand \@ifnum [1]{%
 \ifnum #1\expandafter \@firstoftwo
 \else \expandafter \@secondoftwo
 \fi
}%
\providecommand \@ifx [1]{%
 \ifx #1\expandafter \@firstoftwo
 \else \expandafter \@secondoftwo
 \fi
}%
\providecommand \natexlab [1]{#1}%
\providecommand \enquote  [1]{``#1''}%
\providecommand \bibnamefont  [1]{#1}%
\providecommand \bibfnamefont [1]{#1}%
\providecommand \citenamefont [1]{#1}%
\providecommand \href@noop [0]{\@secondoftwo}%
\providecommand \href [0]{\begingroup \@sanitize@url \@href}%
\providecommand \@href[1]{\@@startlink{#1}\@@href}%
\providecommand \@@href[1]{\endgroup#1\@@endlink}%
\providecommand \@sanitize@url [0]{\catcode `\\12\catcode `\$12\catcode
  `\&12\catcode `\#12\catcode `\^12\catcode `\_12\catcode `\%12\relax}%
\providecommand \@@startlink[1]{}%
\providecommand \@@endlink[0]{}%
\providecommand \url  [0]{\begingroup\@sanitize@url \@url }%
\providecommand \@url [1]{\endgroup\@href {#1}{\urlprefix }}%
\providecommand \urlprefix  [0]{URL }%
\providecommand \Eprint [0]{\href }%
\providecommand \doibase [0]{http://dx.doi.org/}%
\providecommand \selectlanguage [0]{\@gobble}%
\providecommand \bibinfo  [0]{\@secondoftwo}%
\providecommand \bibfield  [0]{\@secondoftwo}%
\providecommand \translation [1]{[#1]}%
\providecommand \BibitemOpen [0]{}%
\providecommand \bibitemStop [0]{}%
\providecommand \bibitemNoStop [0]{.\EOS\space}%
\providecommand \EOS [0]{\spacefactor3000\relax}%
\providecommand \BibitemShut  [1]{\csname bibitem#1\endcsname}%
\let\auto@bib@innerbib\@empty
%</preamble>
\bibitem [{\citenamefont {Lee}\ and\ \citenamefont
  {Sadeghpour}(2013)}]{lee_prl}%
  \BibitemOpen
  \bibfield  {author} {\bibinfo {author} {\bibfnamefont {T.~E.}\ \bibnamefont
  {Lee}}\ and\ \bibinfo {author} {\bibfnamefont {H.~R.}\ \bibnamefont
  {Sadeghpour}},\ }\bibfield  {title} {\enquote {\bibinfo {title} {Quantum
  synchronization of quantum van der pol oscillators with trapped ions},}\
  }\href@noop {} {\bibfield  {journal} {\bibinfo  {journal} {Phys. Rev. Lett.}\
  }\textbf {\bibinfo {volume} {111}},\ \bibinfo {pages} {234101} (\bibinfo
  {year} {2013})}\BibitemShut {NoStop}%
\bibitem [{\citenamefont {Walter}, \citenamefont {Nunnenkamp},\ and\
  \citenamefont {Bruder}(2014)}]{brud_prl1}%
  \BibitemOpen
  \bibfield  {author} {\bibinfo {author} {\bibfnamefont {S.}~\bibnamefont
  {Walter}}, \bibinfo {author} {\bibfnamefont {A.}~\bibnamefont {Nunnenkamp}},
  \ and\ \bibinfo {author} {\bibfnamefont {C.}~\bibnamefont {Bruder}},\
  }\bibfield  {title} {\enquote {\bibinfo {title} {Quantum synchronization of a
  driven self-sustained oscillator},}\ }\href@noop {} {\bibfield  {journal}
  {\bibinfo  {journal} {Phys. Rev. Lett.}\ }\textbf {\bibinfo {volume} {112}},\
  \bibinfo {pages} {094102} (\bibinfo {year} {2014})}\BibitemShut {NoStop}%
\bibitem [{\citenamefont {Laskar}\ \emph {et~al.}(2020)\citenamefont {Laskar},
  \citenamefont {Adhikary}, \citenamefont {Mondal}, \citenamefont {Katiyar},
  \citenamefont {Vinjanampathy},\ and\ \citenamefont {Ghosh}}]{expt1}%
  \BibitemOpen
  \bibfield  {author} {\bibinfo {author} {\bibfnamefont {A.~W.}\ \bibnamefont
  {Laskar}}, \bibinfo {author} {\bibfnamefont {P.}~\bibnamefont {Adhikary}},
  \bibinfo {author} {\bibfnamefont {S.}~\bibnamefont {Mondal}}, \bibinfo
  {author} {\bibfnamefont {P.}~\bibnamefont {Katiyar}}, \bibinfo {author}
  {\bibfnamefont {S.}~\bibnamefont {Vinjanampathy}}, \ and\ \bibinfo {author}
  {\bibfnamefont {S.}~\bibnamefont {Ghosh}},\ }\bibfield  {title} {\enquote
  {\bibinfo {title} {Observation of quantum phase synchronization in spin-1
  atoms},}\ }\href@noop {} {\bibfield  {journal} {\bibinfo  {journal} {Phys.
  Rev. Lett.}\ }\textbf {\bibinfo {volume} {125}},\ \bibinfo {pages} {013601}
  (\bibinfo {year} {2020})}\BibitemShut {NoStop}%
\bibitem [{\citenamefont {Koppenh{\"{o}}fer}, \citenamefont {Bruder},\ and\
  \citenamefont {Roulet}(2020)}]{expt2}%
  \BibitemOpen
  \bibfield  {author} {\bibinfo {author} {\bibfnamefont {M.}~\bibnamefont
  {Koppenh{\"{o}}fer}}, \bibinfo {author} {\bibfnamefont {C.}~\bibnamefont
  {Bruder}}, \ and\ \bibinfo {author} {\bibfnamefont {A.}~\bibnamefont
  {Roulet}},\ }\bibfield  {title} {\enquote {\bibinfo {title} {Quantum
  synchronization on the {IBM Q} system},}\ }\href@noop {} {\bibfield
  {journal} {\bibinfo  {journal} {Phys. Rev. Research}\ }\textbf {\bibinfo
  {volume} {2}},\ \bibinfo {pages} {023026} (\bibinfo {year}
  {2020})}\BibitemShut {NoStop}%
\bibitem [{\citenamefont {Pikovsky}, \citenamefont {Rosenblum},\ and\
  \citenamefont {Kurths}(2003)}]{piko}%
  \BibitemOpen
  \bibfield  {author} {\bibinfo {author} {\bibfnamefont {A.}~\bibnamefont
  {Pikovsky}}, \bibinfo {author} {\bibfnamefont {M.}~\bibnamefont {Rosenblum}},
  \ and\ \bibinfo {author} {\bibfnamefont {J.}~\bibnamefont {Kurths}},\
  }\href@noop {} {\emph {\bibinfo {title} {Synchronization: A Universal Concept
  in Nonlinear Sciences}}}\ (\bibinfo  {publisher} {Cambridge University Press,
  England},\ \bibinfo {year} {2003})\BibitemShut {NoStop}%
\bibitem [{\citenamefont {Carmichael}(1999)}]{carmichael}%
  \BibitemOpen
  \bibfield  {author} {\bibinfo {author} {\bibfnamefont {H.~J.}\ \bibnamefont
  {Carmichael}},\ }\href@noop {} {\emph {\bibinfo {title} {Statistical Methods
  in Quantum Optics 1}}}\ (\bibinfo  {publisher} {Springer},\ \bibinfo {year}
  {1999})\BibitemShut {NoStop}%
\bibitem [{\citenamefont {Weinbub}\ and\ \citenamefont {Ferry}(2018)}]{wigner}%
  \BibitemOpen
  \bibfield  {author} {\bibinfo {author} {\bibfnamefont {J.}~\bibnamefont
  {Weinbub}}\ and\ \bibinfo {author} {\bibfnamefont {D.~K.}\ \bibnamefont
  {Ferry}},\ }\bibfield  {title} {\enquote {\bibinfo {title} {Recent advances
  in wigner function approaches},}\ }\href@noop {} {\bibfield  {journal}
  {\bibinfo  {journal} {Appl. Phys. Rev.}\ }\textbf {\bibinfo {volume} {5}},\
  \bibinfo {pages} {041104} (\bibinfo {year} {2018})}\BibitemShut {NoStop}%
\bibitem [{\citenamefont {Lee}, \citenamefont {Chan},\ and\ \citenamefont
  {Wang}(2014)}]{lee_pre_1}%
  \BibitemOpen
  \bibfield  {author} {\bibinfo {author} {\bibfnamefont {T.~E.}\ \bibnamefont
  {Lee}}, \bibinfo {author} {\bibfnamefont {C.-K.}\ \bibnamefont {Chan}}, \
  and\ \bibinfo {author} {\bibfnamefont {S.}~\bibnamefont {Wang}},\ }\bibfield
  {title} {\enquote {\bibinfo {title} {Entanglement tongue and quantum
  synchronization of disordered oscillators},}\ }\href@noop {} {\bibfield
  {journal} {\bibinfo  {journal} {Phys. Rev. E}\ }\textbf {\bibinfo {volume}
  {89}},\ \bibinfo {pages} {022913} (\bibinfo {year} {2014})}\BibitemShut
  {NoStop}%
\bibitem [{\citenamefont {Walter}, \citenamefont {Nunnenkamp},\ and\
  \citenamefont {Bruder}(2015)}]{brud-ann15}%
  \BibitemOpen
  \bibfield  {author} {\bibinfo {author} {\bibfnamefont {S.}~\bibnamefont
  {Walter}}, \bibinfo {author} {\bibfnamefont {A.}~\bibnamefont {Nunnenkamp}},
  \ and\ \bibinfo {author} {\bibfnamefont {C.}~\bibnamefont {Bruder}},\
  }\bibfield  {title} {\enquote {\bibinfo {title} {Quantum synchronization of
  two van der pol oscillators},}\ }\href@noop {} {\bibfield  {journal}
  {\bibinfo  {journal} {Ann. der. Phys.}\ }\textbf {\bibinfo {volume} {527}},\
  \bibinfo {pages} {131} (\bibinfo {year} {2015})}\BibitemShut {NoStop}%
\bibitem [{\citenamefont {L{\"{o}}rch}\ \emph {et~al.}(2017)\citenamefont
  {L{\"{o}}rch}, \citenamefont {Nigg}, \citenamefont {Nunnenkamp},
  \citenamefont {Tiwari},\ and\ \citenamefont {Bruder}}]{brudprl_17}%
  \BibitemOpen
  \bibfield  {author} {\bibinfo {author} {\bibfnamefont {N.}~\bibnamefont
  {L{\"{o}}rch}}, \bibinfo {author} {\bibfnamefont {S.~E.}\ \bibnamefont
  {Nigg}}, \bibinfo {author} {\bibfnamefont {A.}~\bibnamefont {Nunnenkamp}},
  \bibinfo {author} {\bibfnamefont {R.~P.}\ \bibnamefont {Tiwari}}, \ and\
  \bibinfo {author} {\bibfnamefont {C.}~\bibnamefont {Bruder}},\ }\bibfield
  {title} {\enquote {\bibinfo {title} {Quantum synchronization blockade: Energy
  quantization hinders synchronization of identical oscillators},}\ }\href@noop
  {} {\bibfield  {journal} {\bibinfo  {journal} {Phys. Rev. Lett.}\ }\textbf
  {\bibinfo {volume} {118}},\ \bibinfo {pages} {243602} (\bibinfo {year}
  {2017})}\BibitemShut {NoStop}%
\bibitem [{\citenamefont {Morgan}\ and\ \citenamefont
  {Hinrichsen}(2015)}]{morgan}%
  \BibitemOpen
  \bibfield  {author} {\bibinfo {author} {\bibfnamefont {L.}~\bibnamefont
  {Morgan}}\ and\ \bibinfo {author} {\bibfnamefont {H.}~\bibnamefont
  {Hinrichsen}},\ }\bibfield  {title} {\enquote {\bibinfo {title} {Oscillation
  and synchronization of two quantum self-sustained oscillators},}\ }\href@noop
  {} {\bibfield  {journal} {\bibinfo  {journal} {J. Stat. Mech.}\ }\textbf
  {\bibinfo {volume} {28}},\ \bibinfo {pages} {P09009} (\bibinfo {year}
  {2015})}\BibitemShut {NoStop}%
\bibitem [{\citenamefont {Sonar}\ \emph {et~al.}(2018)\citenamefont {Sonar},
  \citenamefont {Hajdu{\v{s}}ek}, \citenamefont {Mukherjee}, \citenamefont
  {Fazio}, \citenamefont {Vedral}, \citenamefont {Vinjanampathy},\ and\
  \citenamefont {Kwek}}]{squeezing}%
  \BibitemOpen
  \bibfield  {author} {\bibinfo {author} {\bibfnamefont {S.}~\bibnamefont
  {Sonar}}, \bibinfo {author} {\bibfnamefont {M.}~\bibnamefont
  {Hajdu{\v{s}}ek}}, \bibinfo {author} {\bibfnamefont {M.}~\bibnamefont
  {Mukherjee}}, \bibinfo {author} {\bibfnamefont {R.}~\bibnamefont {Fazio}},
  \bibinfo {author} {\bibfnamefont {V.}~\bibnamefont {Vedral}}, \bibinfo
  {author} {\bibfnamefont {S.}~\bibnamefont {Vinjanampathy}}, \ and\ \bibinfo
  {author} {\bibfnamefont {L.}~\bibnamefont {Kwek}},\ }\bibfield  {title}
  {\enquote {\bibinfo {title} {Squeezing enhances quantum synchronization},}\
  }\href@noop {} {\bibfield  {journal} {\bibinfo  {journal} {Phys. Rev. Lett.}\
  }\textbf {\bibinfo {volume} {120}},\ \bibinfo {pages} {163601} (\bibinfo
  {year} {2018})}\BibitemShut {NoStop}%
\bibitem [{\citenamefont {Mok}, \citenamefont {Kwek},\ and\ \citenamefont
  {Heimonen}(2020)}]{enhance-kwek}%
  \BibitemOpen
  \bibfield  {author} {\bibinfo {author} {\bibfnamefont {W.-K.}\ \bibnamefont
  {Mok}}, \bibinfo {author} {\bibfnamefont {L.-C.}\ \bibnamefont {Kwek}}, \
  and\ \bibinfo {author} {\bibfnamefont {H.}~\bibnamefont {Heimonen}},\
  }\bibfield  {title} {\enquote {\bibinfo {title} {Synchronization boost with
  single-photon dissipation in the deep quantum regime},}\ }\href@noop {}
  {\bibfield  {journal} {\bibinfo  {journal} {Phys. Rev. Res.}\ }\textbf
  {\bibinfo {volume} {2}},\ \bibinfo {pages} {033422} (\bibinfo {year}
  {2020})}\BibitemShut {NoStop}%
\bibitem [{\citenamefont {Zakharova}(2020)}]{annabook}%
  \BibitemOpen
  \bibfield  {author} {\bibinfo {author} {\bibfnamefont {A.}~\bibnamefont
  {Zakharova}},\ }\href@noop {} {\emph {\bibinfo {title} {Chimera Patterns in
  Networks}}}\ (\bibinfo  {publisher} {Springer},\ \bibinfo {address} {Cham},\
  \bibinfo {year} {2020})\BibitemShut {NoStop}%
\bibitem [{\citenamefont {Banerjee}\ \emph
  {et~al.}(2018{\natexlab{a}})\citenamefont {Banerjee}, \citenamefont {Biswas},
  \citenamefont {Ghosh}, \citenamefont {Sch{\"o}ll},\ and\ \citenamefont
  {Zakharova}}]{acamc}%
  \BibitemOpen
  \bibfield  {author} {\bibinfo {author} {\bibfnamefont {T.}~\bibnamefont
  {Banerjee}}, \bibinfo {author} {\bibfnamefont {D.}~\bibnamefont {Biswas}},
  \bibinfo {author} {\bibfnamefont {D.}~\bibnamefont {Ghosh}}, \bibinfo
  {author} {\bibfnamefont {E.}~\bibnamefont {Sch{\"o}ll}}, \ and\ \bibinfo
  {author} {\bibfnamefont {A.}~\bibnamefont {Zakharova}},\ }\bibfield  {title}
  {\enquote {\bibinfo {title} {Networks of coupled oscillators: from phase to
  amplitude chimeras},}\ }\href@noop {} {\bibfield  {journal} {\bibinfo
  {journal} {Chaos}\ }\textbf {\bibinfo {volume} {28}},\ \bibinfo {pages}
  {113124} (\bibinfo {year} {2018}{\natexlab{a}})}\BibitemShut {NoStop}%
\bibitem [{\citenamefont {Poel}, \citenamefont {Zakharova},\ and\ \citenamefont
  {Sch{\"o}ll}(2015)}]{annapdeath}%
  \BibitemOpen
  \bibfield  {author} {\bibinfo {author} {\bibfnamefont {W.}~\bibnamefont
  {Poel}}, \bibinfo {author} {\bibfnamefont {A.}~\bibnamefont {Zakharova}}, \
  and\ \bibinfo {author} {\bibfnamefont {E.}~\bibnamefont {Sch{\"o}ll}},\
  }\bibfield  {title} {\enquote {\bibinfo {title} {Partial synchronization and
  partial amplitude death in mesoscale network motifs},}\ }\href@noop {}
  {\bibfield  {journal} {\bibinfo  {journal} {Phy. Rev. E}\ }\textbf {\bibinfo
  {volume} {91}},\ \bibinfo {pages} {022915} (\bibinfo {year}
  {2015})}\BibitemShut {NoStop}%
\bibitem [{\citenamefont {Bastidas}\ \emph {et~al.}(2015)\citenamefont
  {Bastidas}, \citenamefont {Omelchenko}, \citenamefont {Zakharova},
  \citenamefont {Sch{\"{o}}ll},\ and\ \citenamefont {Brandes}}]{schoell_qm}%
  \BibitemOpen
  \bibfield  {author} {\bibinfo {author} {\bibfnamefont {V.~M.}\ \bibnamefont
  {Bastidas}}, \bibinfo {author} {\bibfnamefont {I.}~\bibnamefont
  {Omelchenko}}, \bibinfo {author} {\bibfnamefont {A.}~\bibnamefont
  {Zakharova}}, \bibinfo {author} {\bibfnamefont {E.}~\bibnamefont
  {Sch{\"{o}}ll}}, \ and\ \bibinfo {author} {\bibfnamefont {T.}~\bibnamefont
  {Brandes}},\ }\bibfield  {title} {\enquote {\bibinfo {title} {Quantum
  signatures of chimera states},}\ }\href@noop {} {\bibfield  {journal}
  {\bibinfo  {journal} {Phys. Rev. E}\ }\textbf {\bibinfo {volume} {92}},\
  \bibinfo {pages} {062924} (\bibinfo {year} {2015})}\BibitemShut {NoStop}%
\bibitem [{\citenamefont {Pyragas}(1992)}]{py92}%
  \BibitemOpen
  \bibfield  {author} {\bibinfo {author} {\bibfnamefont {K.}~\bibnamefont
  {Pyragas}},\ }\bibfield  {title} {\enquote {\bibinfo {title} {Continuous
  control of chaos by self-controlling feedback},}\ }\href@noop {} {\bibfield
  {journal} {\bibinfo  {journal} {Physics Letters A}\ }\textbf {\bibinfo
  {volume} {170}},\ \bibinfo {pages} {421--428} (\bibinfo {year}
  {1992})}\BibitemShut {NoStop}%
\bibitem [{\citenamefont {Droenner}\ \emph {et~al.}(2019)\citenamefont
  {Droenner}, \citenamefont {Naumann}, \citenamefont {Sch{\"{o}}ll},
  \citenamefont {Knorr},\ and\ \citenamefont {Carmele}}]{schoellqm2}%
  \BibitemOpen
  \bibfield  {author} {\bibinfo {author} {\bibfnamefont {L.}~\bibnamefont
  {Droenner}}, \bibinfo {author} {\bibfnamefont {N.~L.}\ \bibnamefont
  {Naumann}}, \bibinfo {author} {\bibfnamefont {E.}~\bibnamefont
  {Sch{\"{o}}ll}}, \bibinfo {author} {\bibfnamefont {A.}~\bibnamefont {Knorr}},
  \ and\ \bibinfo {author} {\bibfnamefont {A.}~\bibnamefont {Carmele}},\
  }\bibfield  {title} {\enquote {\bibinfo {title} {Quantum pyragas control:
  Selective control of individual photon probabilities},}\ }\href@noop {}
  {\bibfield  {journal} {\bibinfo  {journal} {Phys. Rev. A}\ }\textbf {\bibinfo
  {volume} {99}},\ \bibinfo {pages} {023840} (\bibinfo {year}
  {2019})}\BibitemShut {NoStop}%
\bibitem [{\citenamefont {Kato}\ and\ \citenamefont {Nakao}(2021)}]{qmcores}%
  \BibitemOpen
  \bibfield  {author} {\bibinfo {author} {\bibfnamefont {Y.}~\bibnamefont
  {Kato}}\ and\ \bibinfo {author} {\bibfnamefont {H.}~\bibnamefont {Nakao}},\
  }\bibfield  {title} {\enquote {\bibinfo {title} {Quantum coherence
  resonance},}\ }\href@noop {} {\bibfield  {journal} {\bibinfo  {journal} {New
  J. Phys.}\ }\textbf {\bibinfo {volume} {23}},\ \bibinfo {pages} {043018}
  (\bibinfo {year} {2021})}\BibitemShut {NoStop}%
\bibitem [{\citenamefont {Ishibashi}\ and\ \citenamefont
  {Kanamoto}(2017)}]{qad1}%
  \BibitemOpen
  \bibfield  {author} {\bibinfo {author} {\bibfnamefont {K.}~\bibnamefont
  {Ishibashi}}\ and\ \bibinfo {author} {\bibfnamefont {R.}~\bibnamefont
  {Kanamoto}},\ }\bibfield  {title} {\enquote {\bibinfo {title} {Oscillation
  collapse in coupled quantum van der pol oscillators},}\ }\href@noop {}
  {\bibfield  {journal} {\bibinfo  {journal} {Phys. Rev. E}\ }\textbf {\bibinfo
  {volume} {96}},\ \bibinfo {pages} {052210} (\bibinfo {year}
  {2017})}\BibitemShut {NoStop}%
\bibitem [{\citenamefont {Amitai}\ \emph {et~al.}(2018)\citenamefont {Amitai},
  \citenamefont {Koppenh{\"{o}}fer}, \citenamefont {L{\"{o}}rch},\ and\
  \citenamefont {Bruder}}]{qad2}%
  \BibitemOpen
  \bibfield  {author} {\bibinfo {author} {\bibfnamefont {E.}~\bibnamefont
  {Amitai}}, \bibinfo {author} {\bibfnamefont {M.}~\bibnamefont
  {Koppenh{\"{o}}fer}}, \bibinfo {author} {\bibfnamefont {N.}~\bibnamefont
  {L{\"{o}}rch}}, \ and\ \bibinfo {author} {\bibfnamefont {C.}~\bibnamefont
  {Bruder}},\ }\bibfield  {title} {\enquote {\bibinfo {title} {Quantum effects
  in amplitude death of coupled anharmonic self-oscillators},}\ }\href@noop {}
  {\bibfield  {journal} {\bibinfo  {journal} {Phys. Rev. E}\ }\textbf {\bibinfo
  {volume} {97}},\ \bibinfo {pages} {052203} (\bibinfo {year}
  {2018})}\BibitemShut {NoStop}%
\bibitem [{\citenamefont {Koseska}, \citenamefont {Volkov},\ and\ \citenamefont
  {Kurths}(2013{\natexlab{a}})}]{kosprep}%
  \BibitemOpen
  \bibfield  {author} {\bibinfo {author} {\bibfnamefont {A.}~\bibnamefont
  {Koseska}}, \bibinfo {author} {\bibfnamefont {E.}~\bibnamefont {Volkov}}, \
  and\ \bibinfo {author} {\bibfnamefont {J.}~\bibnamefont {Kurths}},\
  }\bibfield  {title} {\enquote {\bibinfo {title} {Oscillation quenching
  mechanisms: Amplitude vs oscillation death},}\ }\href@noop {} {\bibfield
  {journal} {\bibinfo  {journal} {Phys. Reports}\ }\textbf {\bibinfo {volume}
  {531}},\ \bibinfo {pages} {173} (\bibinfo {year}
  {2013}{\natexlab{a}})}\BibitemShut {NoStop}%
\bibitem [{\citenamefont {Koseska}, \citenamefont {Volkov},\ and\ \citenamefont
  {Kurths}(2013{\natexlab{b}})}]{kosprl}%
  \BibitemOpen
  \bibfield  {author} {\bibinfo {author} {\bibfnamefont {A.}~\bibnamefont
  {Koseska}}, \bibinfo {author} {\bibfnamefont {E.}~\bibnamefont {Volkov}}, \
  and\ \bibinfo {author} {\bibfnamefont {J.}~\bibnamefont {Kurths}},\
  }\bibfield  {title} {\enquote {\bibinfo {title} {Transition from amplitude to
  oscillation death via turing bifurcation},}\ }\href@noop {} {\bibfield
  {journal} {\bibinfo  {journal} {Phys. Rev. Lett}\ }\textbf {\bibinfo {volume}
  {111}},\ \bibinfo {pages} {024103} (\bibinfo {year}
  {2013}{\natexlab{b}})}\BibitemShut {NoStop}%
\bibitem [{\citenamefont {Bandyopadhyay}\ \emph {et~al.}(2020)\citenamefont
  {Bandyopadhyay}, \citenamefont {Khatun}, \citenamefont {Biswas},\ and\
  \citenamefont {Banerjee}}]{qmod}%
  \BibitemOpen
  \bibfield  {author} {\bibinfo {author} {\bibfnamefont {B.}~\bibnamefont
  {Bandyopadhyay}}, \bibinfo {author} {\bibfnamefont {T.}~\bibnamefont
  {Khatun}}, \bibinfo {author} {\bibfnamefont {D.}~\bibnamefont {Biswas}}, \
  and\ \bibinfo {author} {\bibfnamefont {T.}~\bibnamefont {Banerjee}},\
  }\bibfield  {title} {\enquote {\bibinfo {title} {Quantum manifestations of
  homogeneous and inhomogeneous oscillation suppression states},}\ }\href@noop
  {} {\bibfield  {journal} {\bibinfo  {journal} {Phys. Rev. E}\ }\textbf
  {\bibinfo {volume} {102}},\ \bibinfo {pages} {062205} (\bibinfo {year}
  {2020})}\BibitemShut {NoStop}%
\bibitem [{\citenamefont {Zou}\ \emph {et~al.}(2015)\citenamefont {Zou},
  \citenamefont {Senthilkumar}, \citenamefont {Nagao}, \citenamefont {Kiss},
  \citenamefont {Tang}, \citenamefont {Koseska}, \citenamefont {Duan},\ and\
  \citenamefont {Kurths}}]{natcom}%
  \BibitemOpen
  \bibfield  {author} {\bibinfo {author} {\bibfnamefont {W.}~\bibnamefont
  {Zou}}, \bibinfo {author} {\bibfnamefont {D.~V.}\ \bibnamefont
  {Senthilkumar}}, \bibinfo {author} {\bibfnamefont {R.}~\bibnamefont {Nagao}},
  \bibinfo {author} {\bibfnamefont {I.~Z.}\ \bibnamefont {Kiss}}, \bibinfo
  {author} {\bibfnamefont {Y.}~\bibnamefont {Tang}}, \bibinfo {author}
  {\bibfnamefont {A.}~\bibnamefont {Koseska}}, \bibinfo {author} {\bibfnamefont
  {J.}~\bibnamefont {Duan}}, \ and\ \bibinfo {author} {\bibfnamefont
  {J.}~\bibnamefont {Kurths}},\ }\bibfield  {title} {\enquote {\bibinfo {title}
  {Restoration of rhythmicity in diffusively coupled dynamical networks},}\
  }\href@noop {} {\bibfield  {journal} {\bibinfo  {journal} {Nat. Commun.}\
  }\textbf {\bibinfo {volume} {6}},\ \bibinfo {pages} {7709} (\bibinfo {year}
  {2015})}\BibitemShut {NoStop}%
\bibitem [{\citenamefont {Ghosh}, \citenamefont {Banerjee},\ and\ \citenamefont
  {Kurths}(2015)}]{tanryth}%
  \BibitemOpen
  \bibfield  {author} {\bibinfo {author} {\bibfnamefont {D.}~\bibnamefont
  {Ghosh}}, \bibinfo {author} {\bibfnamefont {T.}~\bibnamefont {Banerjee}}, \
  and\ \bibinfo {author} {\bibfnamefont {J.}~\bibnamefont {Kurths}},\
  }\bibfield  {title} {\enquote {\bibinfo {title} {Revival of oscillation from
  mean-field-induced death: Theory and experiment},}\ }\href@noop {} {\bibfield
   {journal} {\bibinfo  {journal} {Phys. Rev. E}\ }\textbf {\bibinfo {volume}
  {92}},\ \bibinfo {pages} {052908} (\bibinfo {year} {2015})}\BibitemShut
  {NoStop}%
\bibitem [{\citenamefont {Banerjee}(2015)}]{tanCD}%
  \BibitemOpen
  \bibfield  {author} {\bibinfo {author} {\bibfnamefont {T.}~\bibnamefont
  {Banerjee}},\ }\bibfield  {title} {\enquote {\bibinfo {title}
  {Mean-field-diffusion--induced chimera death state},}\ }\href@noop {}
  {\bibfield  {journal} {\bibinfo  {journal} {Europhys. Lett.}\ }\textbf
  {\bibinfo {volume} {110}},\ \bibinfo {pages} {60003} (\bibinfo {year}
  {2015})}\BibitemShut {NoStop}%
\bibitem [{\citenamefont {Banerjee}, \citenamefont {Dutta},\ and\ \citenamefont
  {Gupta}(2015)}]{bandutta}%
  \BibitemOpen
  \bibfield  {author} {\bibinfo {author} {\bibfnamefont {T.}~\bibnamefont
  {Banerjee}}, \bibinfo {author} {\bibfnamefont {P.~S.}\ \bibnamefont {Dutta}},
  \ and\ \bibinfo {author} {\bibfnamefont {A.}~\bibnamefont {Gupta}},\
  }\bibfield  {title} {\enquote {\bibinfo {title} {Mean-field
  dispersion-induced spatial synchrony, oscillation and amplitude death, and
  temporal stability in an ecological model},}\ }\href@noop {} {\bibfield
  {journal} {\bibinfo  {journal} {Phys. Rev. E}\ }\textbf {\bibinfo {volume}
  {91}},\ \bibinfo {pages} {052919} (\bibinfo {year} {2015})}\BibitemShut
  {NoStop}%
\bibitem [{\citenamefont {van~der Pol}(1922)}]{vdp}%
  \BibitemOpen
  \bibfield  {author} {\bibinfo {author} {\bibfnamefont {B.}~\bibnamefont
  {van~der Pol}},\ }\bibfield  {title} {\enquote {\bibinfo {title} {On
  oscillation hysteresis in a triode generator with two degrees of freedom},}\
  }\href@noop {} {\bibfield  {journal} {\bibinfo  {journal} {Philos. Mag.}\
  }\textbf {\bibinfo {volume} {43}},\ \bibinfo {pages} {700--719} (\bibinfo
  {year} {1922})}\BibitemShut {NoStop}%
\bibitem [{\citenamefont {Johansson}, \citenamefont {Nation},\ and\
  \citenamefont {Nori}(2013)}]{qutip}%
  \BibitemOpen
  \bibfield  {author} {\bibinfo {author} {\bibfnamefont {J.}~\bibnamefont
  {Johansson}}, \bibinfo {author} {\bibfnamefont {P.}~\bibnamefont {Nation}}, \
  and\ \bibinfo {author} {\bibfnamefont {F.}~\bibnamefont {Nori}},\ }\bibfield
  {title} {\enquote {\bibinfo {title} {Qutip 2: A python framework for the
  dynamics of open quantum systems},}\ }\href@noop {} {\bibfield  {journal}
  {\bibinfo  {journal} {Comput. Phys. Commun.}\ }\textbf {\bibinfo {volume}
  {184}},\ \bibinfo {pages} {1234} (\bibinfo {year} {2013})}\BibitemShut
  {NoStop}%
\bibitem [{\citenamefont {Ullner}\ \emph {et~al.}(2007)\citenamefont {Ullner},
  \citenamefont {Zaikin}, \citenamefont {Volkov},\ and\ \citenamefont
  {Garc{\'{i}}a-Ojalvo}}]{qstr2}%
  \BibitemOpen
  \bibfield  {author} {\bibinfo {author} {\bibfnamefont {E.}~\bibnamefont
  {Ullner}}, \bibinfo {author} {\bibfnamefont {A.}~\bibnamefont {Zaikin}},
  \bibinfo {author} {\bibfnamefont {E.~I.}\ \bibnamefont {Volkov}}, \ and\
  \bibinfo {author} {\bibfnamefont {J.}~\bibnamefont {Garc{\'{i}}a-Ojalvo}},\
  }\bibfield  {title} {\enquote {\bibinfo {title} {Multistability and
  clustering in a population of synthetic genetic oscillators via
  phase-repulsive cell-to-cell communication},}\ }\href@noop {} {\bibfield
  {journal} {\bibinfo  {journal} {Phys. Rev. Lett}\ }\textbf {\bibinfo {volume}
  {99}},\ \bibinfo {pages} {148103} (\bibinfo {year} {2007})}\BibitemShut
  {NoStop}%
\bibitem [{\citenamefont {Banerjee}\ and\ \citenamefont
  {Ghosh}(2014{\natexlab{a}})}]{tanpre1}%
  \BibitemOpen
  \bibfield  {author} {\bibinfo {author} {\bibfnamefont {T.}~\bibnamefont
  {Banerjee}}\ and\ \bibinfo {author} {\bibfnamefont {D.}~\bibnamefont
  {Ghosh}},\ }\bibfield  {title} {\enquote {\bibinfo {title} {Transition from
  amplitude to oscillation death under mean-field diffusive coupling},}\
  }\href@noop {} {\bibfield  {journal} {\bibinfo  {journal} {Phys. Rev. E}\
  }\textbf {\bibinfo {volume} {89}},\ \bibinfo {pages} {052912} (\bibinfo
  {year} {2014}{\natexlab{a}})}\BibitemShut {NoStop}%
\bibitem [{\citenamefont {Banerjee}\ and\ \citenamefont
  {Ghosh}(2014{\natexlab{b}})}]{tanpre2}%
  \BibitemOpen
  \bibfield  {author} {\bibinfo {author} {\bibfnamefont {T.}~\bibnamefont
  {Banerjee}}\ and\ \bibinfo {author} {\bibfnamefont {D.}~\bibnamefont
  {Ghosh}},\ }\bibfield  {title} {\enquote {\bibinfo {title} {Experimental
  observation of a transition from amplitude to oscillation death in coupled
  oscillators},}\ }\href@noop {} {\bibfield  {journal} {\bibinfo  {journal}
  {Phys. Rev. E}\ }\textbf {\bibinfo {volume} {89}},\ \bibinfo {pages} {062902}
  (\bibinfo {year} {2014}{\natexlab{b}})}\BibitemShut {NoStop}%
\bibitem [{\citenamefont {Ansmann}(2018)}]{jitcode}%
  \BibitemOpen
  \bibfield  {author} {\bibinfo {author} {\bibfnamefont {G.}~\bibnamefont
  {Ansmann}},\ }\bibfield  {title} {\enquote {\bibinfo {title} {Efficiently and
  easily integrating differential equations with {JiTCODE}, {JiTCDDE}, and
  {JiTCSDE}},}\ }\href@noop {} {\bibfield  {journal} {\bibinfo  {journal}
  {Chaos}\ }\textbf {\bibinfo {volume} {28}},\ \bibinfo {pages} {043116}
  (\bibinfo {year} {2018})}\BibitemShut {NoStop}%
\bibitem [{\citenamefont {Zou}, \citenamefont {Zhan},\ and\ \citenamefont
  {Kurths}(2017)}]{lpf}%
  \BibitemOpen
  \bibfield  {author} {\bibinfo {author} {\bibfnamefont {W.}~\bibnamefont
  {Zou}}, \bibinfo {author} {\bibfnamefont {M.}~\bibnamefont {Zhan}}, \ and\
  \bibinfo {author} {\bibfnamefont {J.}~\bibnamefont {Kurths}},\ }\bibfield
  {title} {\enquote {\bibinfo {title} {Revoking amplitude and oscillation
  deaths by low-pass filter in coupled oscillators},}\ }\href@noop {}
  {\bibfield  {journal} {\bibinfo  {journal} {Phy. Rev. E}\ }\textbf {\bibinfo
  {volume} {95}},\ \bibinfo {pages} {062206} (\bibinfo {year}
  {2017})}\BibitemShut {NoStop}%
\bibitem [{\citenamefont {Banerjee}\ \emph
  {et~al.}(2018{\natexlab{b}})\citenamefont {Banerjee}, \citenamefont {Biswas},
  \citenamefont {Ghosh}, \citenamefont {Bandyopadhyay},\ and\ \citenamefont
  {Kurths}}]{banihlc}%
  \BibitemOpen
  \bibfield  {author} {\bibinfo {author} {\bibfnamefont {T.}~\bibnamefont
  {Banerjee}}, \bibinfo {author} {\bibfnamefont {D.}~\bibnamefont {Biswas}},
  \bibinfo {author} {\bibfnamefont {D.}~\bibnamefont {Ghosh}}, \bibinfo
  {author} {\bibfnamefont {B.}~\bibnamefont {Bandyopadhyay}}, \ and\ \bibinfo
  {author} {\bibfnamefont {J.}~\bibnamefont {Kurths}},\ }\bibfield  {title}
  {\enquote {\bibinfo {title} {Transition from homogeneous to inhomogeneous
  limit cycles: Effect of local filtering in coupled oscillators},}\
  }\href@noop {} {\bibfield  {journal} {\bibinfo  {journal} {Phys. Rev. E}\
  }\textbf {\bibinfo {volume} {97}},\ \bibinfo {pages} {042218} (\bibinfo
  {year} {2018}{\natexlab{b}})}\BibitemShut {NoStop}%
\bibitem [{\citenamefont {Zou}\ \emph {et~al.}(2013)\citenamefont {Zou},
  \citenamefont {Senthilkumar}, \citenamefont {Zhan},\ and\ \citenamefont
  {Kurths}}]{dvprl}%
  \BibitemOpen
  \bibfield  {author} {\bibinfo {author} {\bibfnamefont {W.}~\bibnamefont
  {Zou}}, \bibinfo {author} {\bibfnamefont {D.}~\bibnamefont {Senthilkumar}},
  \bibinfo {author} {\bibfnamefont {M.}~\bibnamefont {Zhan}}, \ and\ \bibinfo
  {author} {\bibfnamefont {J.}~\bibnamefont {Kurths}},\ }\bibfield  {title}
  {\enquote {\bibinfo {title} {Reviving oscillations in coupled nonlinear
  oscillators},}\ }\href@noop {} {\bibfield  {journal} {\bibinfo  {journal}
  {Phys. Rev. Lett}\ }\textbf {\bibinfo {volume} {111}},\ \bibinfo {pages}
  {014101} (\bibinfo {year} {2013})}\BibitemShut {NoStop}%
\bibitem [{\citenamefont {Biswas}\ and\ \citenamefont
  {Banerjee}(2018)}]{tbook}%
  \BibitemOpen
  \bibfield  {author} {\bibinfo {author} {\bibfnamefont {D.}~\bibnamefont
  {Biswas}}\ and\ \bibinfo {author} {\bibfnamefont {T.}~\bibnamefont
  {Banerjee}},\ }\href@noop {} {\emph {\bibinfo {title} {Time-Delayed Chaotic
  Dynamical Systems}}}\ (\bibinfo  {publisher} {Springer International
  Publishing},\ \bibinfo {year} {2018})\BibitemShut {NoStop}%
\bibitem [{\citenamefont {Bemani}\ \emph {et~al.}(2017)\citenamefont {Bemani},
  \citenamefont {Motazedifard}, \citenamefont {Roknizadeh}, \citenamefont
  {Naderi},\ and\ \citenamefont {Vitali}}]{21_bemani}%
  \BibitemOpen
  \bibfield  {author} {\bibinfo {author} {\bibfnamefont {F.}~\bibnamefont
  {Bemani}}, \bibinfo {author} {\bibfnamefont {A.}~\bibnamefont
  {Motazedifard}}, \bibinfo {author} {\bibfnamefont {R.}~\bibnamefont
  {Roknizadeh}}, \bibinfo {author} {\bibfnamefont {M.~H.}\ \bibnamefont
  {Naderi}}, \ and\ \bibinfo {author} {\bibfnamefont {D.}~\bibnamefont
  {Vitali}},\ }\bibfield  {title} {\enquote {\bibinfo {title} {Synchronization
  dynamics of two nanomechanical membranes within a fabry-perot cavity},}\
  }\href@noop {} {\bibfield  {journal} {\bibinfo  {journal} {Phys. Rev. A}\
  }\textbf {\bibinfo {volume} {96}},\ \bibinfo {pages} {023805} (\bibinfo
  {year} {2017})}\BibitemShut {NoStop}%
\bibitem [{\citenamefont {Shim}, \citenamefont {Imboden},\ and\ \citenamefont
  {Mohanty}(2007)}]{22_shim}%
  \BibitemOpen
  \bibfield  {author} {\bibinfo {author} {\bibfnamefont {S.-B.}\ \bibnamefont
  {Shim}}, \bibinfo {author} {\bibfnamefont {M.}~\bibnamefont {Imboden}}, \
  and\ \bibinfo {author} {\bibfnamefont {P.}~\bibnamefont {Mohanty}},\
  }\bibfield  {title} {\enquote {\bibinfo {title} {Synchronized oscillation in
  coupled nanomechanical oscillators},}\ }\href@noop {} {\bibfield  {journal}
  {\bibinfo  {journal} {Science}\ }\textbf {\bibinfo {volume} {316}},\ \bibinfo
  {pages} {95} (\bibinfo {year} {2007})}\BibitemShut {NoStop}%
\bibitem [{\citenamefont {Jayich}\ \emph {et~al.}(2008)\citenamefont {Jayich},
  \citenamefont {Sankey}, \citenamefont {Zwickl}, \citenamefont {Yang},
  \citenamefont {Thompson}, \citenamefont {Girvin}, \citenamefont {Clerk},
  \citenamefont {Marquardt},\ and\ \citenamefont {Harris}}]{expt-mem}%
  \BibitemOpen
  \bibfield  {author} {\bibinfo {author} {\bibfnamefont {A.}~\bibnamefont
  {Jayich}}, \bibinfo {author} {\bibfnamefont {J.}~\bibnamefont {Sankey}},
  \bibinfo {author} {\bibfnamefont {B.}~\bibnamefont {Zwickl}}, \bibinfo
  {author} {\bibfnamefont {C.}~\bibnamefont {Yang}}, \bibinfo {author}
  {\bibfnamefont {J.}~\bibnamefont {Thompson}}, \bibinfo {author}
  {\bibfnamefont {S.}~\bibnamefont {Girvin}}, \bibinfo {author} {\bibfnamefont
  {A.}~\bibnamefont {Clerk}}, \bibinfo {author} {\bibfnamefont
  {F.}~\bibnamefont {Marquardt}}, \ and\ \bibinfo {author} {\bibfnamefont
  {J.}~\bibnamefont {Harris}},\ }\bibfield  {title} {\enquote {\bibinfo {title}
  {Dispersive optomechanics: a membrane inside a cavity},}\ }\href@noop {}
  {\bibfield  {journal} {\bibinfo  {journal} {New J. Phys.}\ }\textbf {\bibinfo
  {volume} {8}},\ \bibinfo {pages} {095008} (\bibinfo {year}
  {2008})}\BibitemShut {NoStop}%
\bibitem [{\citenamefont {Lloyd}\ and\ \citenamefont
  {Braunstein}(1999)}]{sq-app2}%
  \BibitemOpen
  \bibfield  {author} {\bibinfo {author} {\bibfnamefont {S.}~\bibnamefont
  {Lloyd}}\ and\ \bibinfo {author} {\bibfnamefont {S.~L.}\ \bibnamefont
  {Braunstein}},\ }\bibfield  {title} {\enquote {\bibinfo {title} {Quantum
  computation over continuous variables},}\ }\href@noop {} {\bibfield
  {journal} {\bibinfo  {journal} {Phys. Rev. Lett.}\ }\textbf {\bibinfo
  {volume} {82}},\ \bibinfo {pages} {1784--1787} (\bibinfo {year}
  {1999})}\BibitemShut {NoStop}%
\bibitem [{\citenamefont {Goda}\ \emph {et~al.}(2008)\citenamefont {Goda},
  \citenamefont {Mikhailov}, \citenamefont {Saraf}, \citenamefont {Adhikari},
  \citenamefont {McKenzie}, \citenamefont {Ward}, \citenamefont {Vass},
  \citenamefont {Weinstein},\ and\ \citenamefont {Mavalvala}}]{sq-app3}%
  \BibitemOpen
  \bibfield  {author} {\bibinfo {author} {\bibfnamefont {K.}~\bibnamefont
  {Goda}}, \bibinfo {author} {\bibfnamefont {O.~M. E.~E.}\ \bibnamefont
  {Mikhailov}}, \bibinfo {author} {\bibfnamefont {S.}~\bibnamefont {Saraf}},
  \bibinfo {author} {\bibfnamefont {R.}~\bibnamefont {Adhikari}}, \bibinfo
  {author} {\bibfnamefont {K.}~\bibnamefont {McKenzie}}, \bibinfo {author}
  {\bibfnamefont {R.}~\bibnamefont {Ward}}, \bibinfo {author} {\bibfnamefont
  {S.}~\bibnamefont {Vass}}, \bibinfo {author} {\bibfnamefont {A.~J.}\
  \bibnamefont {Weinstein}}, \ and\ \bibinfo {author} {\bibfnamefont
  {N.}~\bibnamefont {Mavalvala}},\ }\bibfield  {title} {\enquote {\bibinfo
  {title} {A quantum-enhanced prototype gravitational-wave detector},}\
  }\href@noop {} {\bibfield  {journal} {\bibinfo  {journal} {Nat. Phys.}\
  }\textbf {\bibinfo {volume} {4}},\ \bibinfo {pages} {472--476} (\bibinfo
  {year} {2008})}\BibitemShut {NoStop}%
\end{thebibliography}
\providecommand{\noopsort}[1]{}\providecommand{\singleletter}[1]{#1}%
\end{document}